\documentclass[12pt]{article}
\usepackage{tocloft}
\usepackage{amsfonts}
\usepackage{amsmath}
\usepackage{amssymb}
\usepackage{array,slashed}
\usepackage{bigints}
\usepackage{bm}
\usepackage{booktabs}
\usepackage[nosort]{cite}
\usepackage{color}
\usepackage{dsfont}
\usepackage{float}
\usepackage{framed}
\usepackage{graphicx}
\usepackage{indentfirst}
\usepackage{mathrsfs}
\usepackage{multirow}
\usepackage{pdflscape}
\usepackage{tikz-cd}
\usepackage{setspace}
\usepackage{subdepth}
\usepackage{subfig}
\usepackage{titlesec}
\usepackage{wrapfig}
\usepackage[all]{xy}
\usepackage{young}
\usepackage[vcentermath]{youngtab}
\usepackage{relsize}
\usepackage{stackengine}
\usepackage{verbatim}
\usepackage{slashed}
\usepackage{color}

\usepackage{hyperref}
\hypersetup{colorlinks=true}
\hypersetup{linkcolor=black}
\hypersetup{citecolor=black}
\hypersetup{urlcolor=black}

\numberwithin{equation}{section}

\usepackage[left=2.5cm,right=2.5cm,top=2.5cm,bottom=3cm]{geometry}
\fontdimen2\font=1.2\fontdimen2{\jot}{5pt}

\titleformat{\section}{\large\bfseries}{\thesection.}{4pt}{}
\titlespacing{\section}{0pt}{20pt}{6pt}

\titleformat{\subsection}{\normalfont\bfseries}{\thesubsection.}{4pt}{}
\titlespacing{\subsection}{0pt}{15pt}{6pt}

\titleformat{\subsubsection}{\normalfont\itshape}{\thesubsubsection.}{4pt}{}
\titlespacing{\subsubsection}{0pt}{15pt}{6pt}

\titleformat{\paragraph}{\normalfont\itshape}{\theparagraph.}{4pt}{}
\titlespacing{\paragraph}{0pt}{15pt}{6pt}

\definecolor{green}{rgb}{0.3,0.7,0.3}

\def\tilde{\widetilde}

\def\hat{\widehat}

\def\bar{\overline}

\DeclareMathAlphabet{\mathbfsf}{OT1}{cmss}{bx}{n}

\def\bZ{\mathbb{Z}}

\newcommand{\cA}{\mathcal{A}}

\newcommand{\cF}{\mathcal{F}}

\newcommand{\cH}{\mathcal{H}}
\newcommand{\cI}{\mathcal{I}}
\newcommand{\cJ}{\mathcal{J}}

\newcommand{\cN}{\mathcal{N}}
\newcommand{\cO}{\mathcal{O}}
\newcommand{\cS}{\mathcal{S}}

\newcommand{\cQ}{\mathcal Q}

\newcommand\scrL{\mathscr{L}}
\newcommand{\ed}{\,.}
\newcommand{\ec}{\,,}

\newcommand{\be}{\begin{equation}}
\newcommand{\ee}{\end{equation}}

\newcommand{\one}{^{(1)}}
\newcommand{\two}{^{(2)}}
\newcommand{\three}{^{(3)}}

\newcommand{\us}{\textrm{us}}

\DeclareFontShape{OT1}{cmr}{mx}{n}%
{<->cmr10}{}
\newcommand{\mytitlefont}{\fontseries{mx}\selectfont}
\DeclareMathAlphabet{\titlemath}{OT1}{cmr}{mx}{n}

\begin{document}
	
	

%
\begin{titlepage}
\begin{flushright} \small
CERN-TH-2026-083
 \end{flushright}
\begin{center}
~\\[20pt]
{\fontsize{27pt}{0pt} \mytitlefont Comments on Symmetry Operators, Asymptotic Charges and Soft Theorems}
\vskip30pt
Luigi Tizzano
\vskip20pt
{~{\it CERN, Theoretical Physics Department, CH-1211 Geneva 23, Switzerland}}%
%
%
\bigskip
\bigskip
\bigskip
\end{center}
\vskip0.2cm
\noindent We study the relation between emergent 1-form symmetries and soft photon theorems in QED. We show that in the relevant massive and massless kinematic regimes, described respectively by HQET and SCET, the soft sector admits electric and magnetic 1-form symmetries. We then show that these symmetries give rise to an infinite-dimensional Abelian algebra of ordinary conserved charges, with a central extension. In Minkowski spacetime, suitable choices of hypersurfaces reduce these charges to the familiar asymptotic symmetry charges and imply the leading electric and magnetic soft photon theorems. We further show that the central term in this algebra fixes a contact term appearing in scattering amplitudes involving two soft photons with mixed electric-magnetic polarizations. Finally, we extend the same construction to inclusive observables and apply it to QED photon detectors.

\vfill 
\begin{flushleft}
April 2026
\end{flushleft}
\end{titlepage}
%
		
	
\setcounter{tocdepth}{3}
\renewcommand{\cfttoctitlefont}{\large\bfseries}
\renewcommand{\cftsecaftersnum}{.}
\renewcommand{\cftsubsecaftersnum}{.}
\renewcommand{\cftsubsubsecaftersnum}{.}
\renewcommand{\cftdotsep}{6}
\renewcommand\contentsname{\centerline{Contents}}
	
\tableofcontents


\section{Introduction and Summary}

The study of scattering amplitudes plays a central role in modern perturbative quantum field theory.
While these amplitudes often suffer from infrared divergences and are not themselves good physical observables, they remain a very useful intermediate step in actual collider physics computations of well-defined observables.

A central insight in this subject is that scattering amplitudes that describe the emission or absorption of soft particles often exhibit a universal behavior and obey a variety of soft theorems. A famous example is the Adler zero \cite{Adler:1964um}, which appears in pion scattering. Another example, which will be of central importance in this paper, is the so-called soft photon theorem \cite{Low:1958sn, Weinberg:1964ew, Weinberg:1965nx, Burnett:1967km}. It states that the leading IR behavior of an $(n+1)$-particle amplitude $\cA_{n+1}$ involving a soft photon of momentum $q$ takes the form
\begin{equation}
   \lim_{q\to0}\cA_{n+1} \to \mathscr{S}\cA_n\ed
\end{equation}
Here $\mathscr{S}$ denotes the soft factor, which depends on the electric
charges and momenta of the hard charged particles, as well as the polarization and momentum of the soft photon. Soft theorems have also been extensively studied over the years in non-Abelian gauge theories and gravity \cite{Weinberg:1964ew, Weinberg:1965nx, Bern:2014vva, Cachazo:2014fwa, Bassetto:1983mvz, Berends:1988zn}. In some favorable cases, soft theorems are robust under quantum corrections, and it can be shown that such corrections are absent or are restricted to a small number of loop orders.\footnote{In this paper, however, we will only discuss tree-level amplitudes.}

The above considerations motivate the question of whether this universality and robustness are rooted in an underlying symmetry principle. One point of view often advocated in the literature relates them to the existence of infinitely many symmetries acting on scattering states at null infinity, known as asymptotic symmetries \cite{Strominger:2013lka, He:2014cra, Campiglia:2015qka, Kapec:2015ena}.\footnote{See \cite{Strominger:2017zoo} for a pedagogical review.} From this perspective, the statement that a scattering amplitude satisfies the Ward identities associated with these asymptotic symmetries is equivalent to the statement that it obeys the corresponding soft theorem.

Recently, \cite{Berean-Dutcher:2025ohp} advocated an alternative point of view. This approach is based on the idea that, in a particular kinematic regime in which particle pair production is suppressed, massive QED exhibits an emergent electric $U(1)\one$ 1-form global symmetry \cite{Gaiotto:2014kfa}. This perspective is particularly natural because the regime of interest is described by a heavy-particle effective theory, in which soft interactions decouple through Wilson lines, the objects that carry charge under the emergent $U(1)_e\one$ 1-form symmetry. One may then invoke the corresponding 1-form Ward identity acting on these soft Wilson lines to derive the soft photon theorem. One virtue of this approach is that it connects more directly with the extensive literature on generalized symmetries (see \cite{Cordova:2022ruw} for a large selection of references), which has significantly broadened our understanding of global symmetries in quantum field theory, but has so far been only sparsely explored in Lorentzian signature,\footnote{See \cite{Casini:2020rgj, Benedetti:2022zbb, Copetti:2024rqj, Harlow:2025cqc} for some exceptions.} where many observables of interest in particle physics are defined.

Motivated by these ideas, our goal in this work is to revisit the leading soft photon theorem in QED and clarify how the above approaches are related. While the emergence of a $U(1)_e\one$ symmetry is elegant, it does not immediately explain why the asymptotic symmetry perspective leads instead to an infinite-dimensional group of symmetries. In that framework, asymptotic symmetries arise from gauge transformations that do not vanish at infinity and therefore act non-trivially on physical states. In the case of the soft photon theorem, these transformations are parametrized by a function $\varepsilon(z,\bar z)$ that is non-vanishing on a two-sphere at the null boundary of Minkowski space. Each such function gives rise to a conserved charge $Q_\varepsilon$, and hence to an infinite family of conserved charges. By contrast, the conserved 1-form symmetry charge identified in \cite{Berean-Dutcher:2025ohp} is unique and does not depend on such data.\footnote{For an early discussion of this point, see \cite{Lake:2018dqm}.} Furthermore, the asymptotic symmetry analysis admits a magnetic refinement of the soft photon theorem \cite{Strominger:2015bla} and applies equally well to both massless and massive QED.

Our starting point is the observation that the EFTs describing the relevant kinematical regimes of QED in the massive and massless limits, formulated respectively using ideas borrowed from heavy-quark effective theory (HQET) \cite{Neubert:1996wg, Manohar:2000dt, Oleari:SCETHQET} and soft-collinear effective theory (SCET) \cite{Bauer:2000yr, Bauer:2001yt}, possess an emergent infinite-dimensional algebra of ordinary 0-form symmetries.\footnote{In the massless case, this statement should be understood as referring to the ultrasoft sector of SCET, where the emergent electric 1-form symmetry is realized, while the collinear sector retains explicit dependence on the gauge potential.
} To make this structure manifest, we adapt the construction of \cite{Hofman:2018lfz}, which describes how such symmetry algebras arise in theories with a $U(1)\one$ 1-form symmetry.\footnote{A recent study of this algebra in the context of the superfluid EFT appeared in \cite{Vitouladitis:2025zoy}.} This infinite-dimensional algebra allows us to reformulate the 1-form symmetry Ward identity in Minkowski space in a way that matches the standard statement found in the asymptotic symmetry literature. The same construction also extends naturally to the magnetic 1-form symmetry $U(1)\one_m$, which, in the absence of dynamical magnetic monopoles, is a conserved global symmetry of QED and similarly implies the leading magnetic soft photon theorem. Finally, the infinite-dimensional algebra we uncover carries a Schwinger term reflecting the mixed anomaly between the electric and magnetic 1-form symmetries. This provides a direct physical application of the anomaly: it fixes the coefficient of the ordering-dependent contact term appearing in scattering amplitudes with soft particles of mixed electric-magnetic helicity.

We further show that the same 1-form symmetry Ward identities extend beyond scattering amplitudes to inclusive observables. This is the natural framework for detector observables, and in particular for the QED photon detector discussed in section \ref{sec:softsq-dglap-qed}, whose leading soft behavior can be understood from the same 1-form symmetry principles. This observable-level perspective is also naturally connected to the language of light-ray operators and  Dokshitzer–Gribov–Lipatov–Altarelli–Parisi (DGLAP) detectors recently discussed in \cite{Chang:2025zib}. The key new ingredient is that inclusive observables are naturally formulated as in--in (``cut'') correlators, so that the symmetry charges act on both sides of the cut. Related questions for inclusive observables in gravity have  been recently studied using asymptotic symmetry methods \cite{Gonzalez:2025ene,Moult:2025njc}.

\subsection{Open Questions}
\begin{itemize}
\item It would be interesting to extend the ideas developed in this work to the IR structure of gravity in asymptotically flat spacetimes and to its BMS symmetry \cite{Bondi:1962px, Sachs:1962wk, Strominger:2013jfa, He:2014laa, Strominger:2014pwa}. Since a soft-collinear effective theory for gravity is already available \cite{Beneke:2012xa, Beneke:2021umj, Beneke:2022pue}, a natural question is whether it realizes the bi-form symmetries studied in \cite{Hinterbichler:2022agn} and whether the corresponding Ward identities reproduce the soft graviton theorem.\footnote{See also \cite{Benedetti:2021lxj, Hull:2024xgo, Cheung:2024ypq, Hull:2024mfb, Hull:2024bcl, Hull:2024ism,  Hull:2024qpy} for other discussions of generalized symmetries in gravity.}
\item Soft theorems are also widely studied in theories with a non-Abelian gauge group $G$. In contrast to QED, even the leading soft gluon theorem receives loop corrections in a non-Abelian theory, and the associated soft divergences are expected to depend in an intricate way on the running coupling and color factors. In \cite{He:2015zea} it was proposed that, at tree level, the soft gluon theorem is equivalent to a Ward identity for a holomorphic $G$ Kac--Moody current algebra.\footnote{See \cite{Magnea:2021fvy, Magnea:2025zut} for some recent works in this direction.} It would be interesting to understand whether this symmetry algebra emerges naturally in the EFT description and how its breaking at the quantum level can be quantified. 
\item There are many long-standing puzzles associated with scattering electrically charged particles off magnetic monopoles. Recent discussions of this problem have emphasized the subtle role of soft dressings, generalized symmetries, and the structure of asymptotic states in electric-magnetic scattering \cite{Csaki:2020inw, Kitano:2021missing, Lippstreu:2021avq, Csaki:2021monopole, Brennan:2021callan, Hamada:2022fock, Csaki:2022dressed, Brennan:2023tae, vanBeest:2023dbu, Mouland:2024zgk}. Since our analysis makes the electric and magnetic 1-form symmetries, as well as their mixed anomaly, manifest in an EFT framework, it is natural to ask whether the contact term derived here is related to the pairwise phases appearing in monopole scattering, and more generally whether the mixed anomaly constrains the infrared structure and crossing properties of electric-magnetic amplitudes.

\end{itemize}

This paper is organized as follows. In section \ref{QED1form}, we review the EFT description of the soft sector of QED in the massive and massless regimes using HQET and SCET, discuss the factorization of soft degrees of freedom in their scattering amplitudes, and explain the role of the (emergent) electric and magnetic 1-form symmetries. In section \ref{symmetriessoftthm}, we show how these 1-form symmetries give rise to an infinite-dimensional algebra of ordinary symmetries and use it to derive the leading electric and magnetic soft photon theorems. We also analyze the Schwinger term associated with the mixed 1-form symmetry anomaly. In section \ref{sec:softsq-dglap-qed}, we extend the same Ward identity logic to inclusive observables and apply it to QED photon detector observables. Appendix \ref{celcoord} reviews our conventions for Bondi coordinates, while appendix \ref{sec:freefield} contains a free-field derivation of the mixed-anomaly contact term.

\section{Emergent 1-Form Symmetry of QED}\label{QED1form}
In this section we develop the effective field theory perspective underlying our analysis of the leading soft photon theorem in QED. The relevant kinematical regimes are described by HQET in the massive case and by SCET in the massless case. In both descriptions, soft interactions are encoded by Wilson lines, making manifest the higher-form symmetry structure of the infrared theory. At the same time, the effective field theory framework makes the factorization of scattering amplitudes into hard, collinear, and soft contributions manifest. 

\subsection{Effective Field Theory Perspective}
\subsubsection{Massive Theory: HQET}
Let us consider massive QED defined by the Lagrangian
\be\label{QED}
\scrL = -\frac{1}{4e^2}F_{\mu\nu}F^{\mu\nu} + \bar{\Psi}(i\slashed{D} - m)\Psi\ec
\ee
where $\Psi$ is a massive Dirac fermion of charge $Q$ and $D_{\mu} = \partial_\mu - iQA_\mu$. We are interested in the regime where the fermion is heavy and nearly on-shell, such that
$p^\mu = m v^\mu + k^\mu\ec$ with four-velocity $v^\mu$ satisfying $v^2=1$ and $k^\mu$ a residual momentum satisfying $|k^\mu|\ll m$. Physically, $k^\mu$ encodes the small off-shell fluctuations of the heavy particle around its mass shell. 

The appropriate framework to describe this limit is Heavy Quark Effective Theory (HQET) \cite{Isgur:1989vq, Isgur:1990yhj, Eichten:1989zv, Georgi:1990um}, (see \cite{Neubert:1996wg, Manohar:2000dt, Oleari:SCETHQET} for pedagogical introductions), obtained by expanding in powers of $k/m$. The first step is to decompose the Dirac field as
\be\label{decomposition}
\Psi(x) = e^{-imv\cdot x}[h_v(x)+H_v(x)]\ec
\ee
where $h_v(x)$ and $H_v(x)$ are respectively the light (particle) and heavy (antiparticle) components, which are obtained by acting with the projectors:
\be
h_v(x) = e^{+im v\cdot x} P_+ \Psi(x)\ec\quad H_v(x) = e^{+im v\cdot x} P_- \Psi(x)\ec\quad P_{\pm} \equiv \frac12(1\pm \gamma^\mu v_\mu)\ed
\ee
Substituting \eqref{decomposition} into the Lagrangian \eqref{QED} we have
\begin{equation}
\scrL = -\frac{1}{4e^2}F_{\mu\nu}F^{\mu\nu} + \bar{h}_v\, i v \cdot D\, h_v - \bar{H}_v (i v \cdot D + 2m) H_v + \left[ \bar{h}_v\, i \slashed{D}_\perp\, H_v + \text{h.c.} \right] \ec
\end{equation}
where, given any tensor $T$, we define its component $T_\perp$ perpendicular to the direction of $v$, by contracting all its indices to the projector 
 $\Pi_\perp^{\mu \nu}=\eta^{\mu \nu}-v^\mu v^\nu$, e.g. $D_\perp^\mu \equiv \Pi_\perp^{\mu \nu} D_\nu$.
 
 Note that $H_v(x)$ has an effective mass gap of $2m$ in the EFT frame and can be integrated out to obtain the following effective theory
\begin{align}
\label{effective1}
\scrL_{\text{HQET}} =&
-\frac{1}{4e^2}F_{\mu\nu}F^{\mu\nu} + i\bar{h}_v\,  (v \cdot D)\, h_v - 
\bar{h}_v\, \slashed{D}_\perp \frac{1}{2m+i v\cdot D} \slashed{D}_\perp  \, h_v 
\\
=&-\frac{1}{4e^2}F_{\mu\nu}F^{\mu\nu} + i\bar{h}_v\,  (v \cdot D)\, h_v - \frac{1}{2m} \bar{h}_v ( D_\perp)^2 h_v + \frac{Q}{4m} \bar{h}_v \sigma^{\mu\nu} F_{\perp \, \mu\nu} h_v + \cdots\ec
\end{align}
where in the second line we define $\sigma^{\mu\nu} = \frac{i}{2}[\gamma^\mu, \gamma^\nu]$ and retain terms through order $O(k/m)$.

An interesting insight from the work \cite{Berean-Dutcher:2025ohp} is that while QED has an explicitly broken electric 1-form symmetry due to pair production—implying that electric flux lines can break by ending on charged particles—the heavy-quark effective theory described above, in which we have integrated out the antiparticles, has no pair production and therefore exhibits an emergent 1-form symmetry protecting flux strings. To make this manifest it is useful to introduce the following change of variables
\be\label{newvar}
h_v(x) = W_Q(C_x)\tilde{h}_v(x)= \exp\left(iQ\int^0_{-\infty} ds  v^\mu A_\mu(x+sv)\right)\tilde{h}_v\ec
\ee
where the trajectory $C_x$ is defined by the curve $y^\mu(s) =x+ sv^\mu$ with $s\in (-\infty,0]$.
 The operator $W_Q(C_x)$ denotes a Wilson line of charge $Q$, which creates a probe of electric charge $Q$ moving along the (timelike) contour $C_x$. These operators will play a fundamental role in this paper. 
Using this change of variables we find
\be
D_\mu h_v= W_Q(C_x) \left[\partial_\mu +\frac{i Q}{v \cdot \partial } F_{\mu \nu} v^{\nu}\right] \tilde{h}_v \equiv W_Q(C_x) L_\mu \tilde{h}_v
\, ,
\ee
where $\frac{1}{v \cdot \partial}F_{\mu \nu}(x)=\int_{-\infty}^0 ds F_{\mu \nu}(x+sv)$, which also implies that the action of covariant derivatives in direction $v$ reduces  to simple derivatives, e.g.
$\bar{h}_v\,  (v \cdot D)\, h_v= \bar{\tilde{h}}_v\,  (v \cdot \partial)\, \tilde{h}_v$.

The effective action in  terms of $\tilde{h}_v$ can thus be written as
\begin{align}
\label{HQET_full}
 \scrL_{\text{HQET}}=
 &-\frac{1}{4e^2} F_{\mu \nu} F^{\mu \nu} +  i\bar{\tilde{h}}_v\,  (v \cdot \partial)\, \tilde{h}_v -
 \bar{\tilde{h}}_v \; L_{\perp}^\mu \frac{\delta_{\mu \nu}-i \frac{\sigma_{\mu \nu}}{2}}{2m+i v\cdot \partial} \; L_{\perp}^\nu  \tilde{h}_v 
 \\
 =  \nonumber
 &-\frac{1}{4e^2}F_{\mu\nu}F^{\mu\nu} + i\bar{\tilde{h}}_v\,  (v \cdot \partial)\, \tilde{h}_v\\ &- \frac{1}{2m} \, \bar{\tilde{h}}_v \left( \partial_j + i Q \int_{-\infty}^0 ds\, v^\nu F_{ j\nu}(x + s v) \right)^2 \tilde{h}_v + \frac{Q}{4m} \bar{\tilde{h}}_v \sigma^{jk} F_{jk} \tilde{h}_v + \cdots\, , 
\end{align}
where the indices $j,k$ label directions perpendicular to $v$; namely, they are projected using $\Pi_\perp$.
 Since the action depends on the gauge field only through $F_{\mu \nu}$,
it is manifestly invariant under the electric $U(1)_e\one$ 1-form symmetry \cite{Gaiotto:2014kfa} acting as
\be\label{1gaugetrans}
A\one \to A\one + \Lambda\one_e\ec\qquad \int_{\Sigma\two} d\Lambda_e\one \in 2\pi \bZ\ec
\ee
where $\Sigma\two$ is a closed 2-cycle. In particular, this is true to all orders in the $k/m$ expansion. 
  Using \eqref{HQET_full} one can also obtain a closed form expression for the associated conserved current  $J_e^{\mu\nu} $,  valid at all orders in $k/m$. It is interesting to notice that $J_e^{\mu\nu} $ is not a local operator, since all the subleading orders in $k/m$ contain operators integrated over the infinite timelike contour $C_x$. 
Nevertheless we are going to be interested mainly in the deep IR properties of the action which are governed by the leading term,  which possesses a true local current equal to $\frac{1}{e^2} F^{\mu\nu}$.
Finally, let us also stress that QED possesses an independent ``magnetic'' 1-form symmetry $U(1)_m^{(1)}$, whose conserved two-form current is $J_m^{\mu\nu} = \tfrac{1}{2\pi} *F^{\mu\nu}$, where $*$ is the Hodge dual operator.\footnote{In Lorentzian signature $*^2=-1$ on 2-forms.} This is an exact symmetry of the theory, broken only by dynamical magnetic monopoles, which are absent in our setup. Consequently, the effective heavy-quark theory~\eqref{HQET_full} also inherits this symmetry.

\subsubsection{Massless Theory: SCET}
Having analyzed the heavy-particle limit of massive QED using HQET in the previous section, we now turn to the opposite kinematic regime relevant for massless QED, where the charged particles are nearly lightlike. Let us consider
\begin{equation} \label{QEDmassless}
\scrL
= -\frac{1}{4e^2} F_{\mu\nu}F^{\mu\nu} + \bar\Psi i\slashed D \Psi\ec
\end{equation}
and final states with energetic, nearly lightlike charged legs with directions $n^\mu_i$ such that $n^2_i=0$. We also introduce a conjugate lightlike vector $\bar{n}_\mu$ such that $n\cdot \bar{n} = 2$ and decompose any 4-vector as
\be\label{decomposed}
p^\mu = \frac{\bar{n}\cdot p}{2}n^\mu + \frac{n\cdot p}{2}\bar{n}^\mu +  p_{\perp}^\mu\ec \qquad n\cdot p_{\perp} = \bar{n}\cdot p_{\perp} =0\ed
\ee
The appropriate effective field theory for describing this regime is the Soft-Collinear Effective Theory (SCET) \cite{Bauer:2000yr, Bauer:2001yt} (see \cite{Becher:2014oda, Oleari:SCETHQET} for pedagogical reviews). Its power counting is organized in terms of a small parameter $\lambda \sim {p_\perp/ \mathcal{Q}}$, which characterizes how collimated or soft degrees of freedom are compared to the large momentum scale $\mathcal{Q}$. With this, one distinguishes modes with scalings
\be
\textrm{collinear}: p^\mu_c \sim \cQ(1,\lambda^2, \lambda)\ec\qquad \textrm{ultrasoft}: p^\mu_{\textrm{us}} \sim \cQ(\lambda^2,\lambda^2,\lambda^2)\ec
\ee
in the components of \eqref{decomposed}. As before, we can introduce a set of (lightcone) projectors
\be
P_n \equiv \frac{(\gamma \cdot \bar{n} ) (\gamma \cdot n)}{4}\ec\qquad P_{\bar{n}} \equiv \frac{(\gamma \cdot n )( \gamma \cdot \bar{n})}{4}\ec
\ee
which we use to split the Dirac field:
\be
\Psi = \xi_n + \eta_n\ec\qquad \xi_n \equiv P_n \Psi\ec \qquad P_{\bar{n}}\Psi \equiv \eta_n\ed
\ee
The fermionic part of the Lagrangian \eqref{QEDmassless} becomes
\be\label{Lferm}
\scrL_f = i\bar{\xi}_n \frac{\gamma \cdot \bar{n}}{2}(n\cdot D)\xi_n - i\bar{\eta}_n \frac{\gamma\cdot n}{2}(\bar{n}\cdot D)\eta_n + (i\bar{\xi}_n \slashed{D}_{\perp} \eta_n + \textrm{h.c.})\ed
\ee
The component $\eta_n$ is off-shell by $\bar n \cdot p \sim \cQ$ therefore it can be integrated out, its associated equation of motion is
\be
\eta_n = \frac{1}{i \bar{n}\cdot D} \frac{\gamma\cdot \bar{n}}{2} i \slashed{D}_\perp \xi_n\ed
\ee
Plugging this back into \eqref{Lferm}, we obtain the leading power collinear Lagrangian
\be
\scrL_{\textrm{c}} = \bar{\xi}_n \frac{\gamma \cdot \bar{n}}{2} \left( i n\cdot D +i\gamma^\mu_\perp D_{\perp \mu}~\frac{1}{\bar{n}\cdot D}\gamma^\nu_\perp D_{\perp\nu}\right)\xi_n\ec
\ee
with $D_\mu = \partial_{\mu} -i Q(A^n_{\mu}+ A^\us_{\mu})$, where $A^n_{\mu}$ is the collinear gauge field which is separated from the ultrasoft gauge field $A^\us_\mu$ by SCET power counting. Also in this context, we introduce the following change of variables
\be\label{xin}
\xi_n(x)=Y_Q(C_{x})\,\tilde\xi_n(x)
=\exp\!\left(iQ\int_{-\infty}^{0} ds\, \frac{n^\mu}{2}\,
A^{\us}_\mu\!\left(x+s\frac{n}{2}\right)\right)\tilde\xi_n(x)\ec
\ee
where the trajectory $C_x$ is defined by the curve $y^\mu(s) = x^\mu +s\frac{n^\mu}{2}$ with $s\in (-\infty,0]$. The Wilson lines $Y_Q(C_x)$ now describe probe electric charges moving on null trajectories. 
\be
\cF_\mu(x)\equiv
i\int_{-\infty}^{0} ds\,\frac{n^\nu}{2}
F^{\us}_{\mu\nu}\!\left(x+s\frac{n}{2}\right)\ed
\ee
With this notation, the $\cO(\lambda^0)$ SCET Lagrangian can be written in a way that mirrors the derivation of HQET:
\be\label{scet0}
\begin{aligned}
\scrL^{(0)}_{\textrm{SCET}}
={}&
-\frac{1}{4e^2}F^n_{\mu\nu}F_n^{\mu\nu} + \bar{\tilde{\xi}}_n\frac{\gamma\cdot\bar n}{2}
\left(
i n\cdot D_c
+i\gamma^\mu_\perp D_{c\perp\mu}
\frac{1}{i\bar n\cdot D_c}
i\gamma^\nu_\perp D_{c\perp\nu}
\right)\tilde{\xi}_n
\\
&-\frac{1}{4e^2}F^\us_{\mu\nu}F_\us^{\mu\nu}
+ \bar\psi_{\us}\,
i\gamma^\mu D_{\us\,\mu}\,
\psi_{\us}~.
\end{aligned}
\ee
where $(D_{c})_\mu=\partial_\mu-iQ A^n_\mu$ and $(D_{\us})_\mu=\partial_\mu-iQ A^\us_\mu$. In close analogy with the HQET construction, the field redefinition above factors out a Wilson line built from the ultrasoft gauge field. There is, however, an important difference from the heavy limit. In HQET the effective action couples to the gauge field only through the field strength $F_{\mu\nu}$ (and nonlocal integrals of $F_{\mu\nu}$ along the heavy worldline). As a result, the full theory is invariant under the electric $U(1)^{(1)}_e$ 1-form symmetry. By contrast, in SCET the collinear modes retain explicit coupling to the collinear gauge potential $A^n_\mu$, and therefore they are not invariant under arbitrary 1-form shifts of that field.

The 1-form symmetry $U(1)\one_e$ can survive only in the ultrasoft sector $A\one_\us \to A\one_\us + \lambda\one_\us$ with the $1$-form $\lambda\one_\us$ subject to the same constraint as in \eqref{1gaugetrans}. However, by direct inspection of the Lagrangian \eqref{scet0}, it is clear that $J\two_e =F\two_{\us}/e^2$ is not conserved in the presence of dynamical ultrasoft charged fermions. Equivalently, ultrasoft Wilson lines can end on dynamical ultrasoft fermion operators. As a result, in the remainder of this work,
our massless analysis is restricted to tree-level amplitudes with no
external ultrasoft fermions. Accordingly, any statement in this work involving the realization or conservation of an electric $U(1)\one_e$ symmetry in SCET should be understood as restricted to this subsector.

Beyond tree level, it is known that connected radiative ultrasoft fermion webs generate an additional leading soft contribution beginning at $\cO(\alpha^3)$ in massless QED
\cite{Ma:2023gir}. These effects lie outside the tree-level
analysis performed here.\footnote{I would like to thank J. Parra-Martinez for pointing out this reference to me.}

Finally, note also that the magnetic 1-form symmetry $U(1)\one_m$ is not affected by the above discussion since (in the absence of dynamical magnetic monopoles) it is a true symmetry of the UV theory.

\subsubsection{1-Form Symmetry Mixed Anomaly}
An important feature of systems with electric and magnetic 1-form symmetries is that they exhibit a mixed 't Hooft anomaly. This is very well-known and we will be brief just to highlight the points that are useful for the present discussion. For more details, see \cite{Gaiotto:2014kfa}; for a particle-physicist-friendly review of the subject, see \cite{Brennan:2023mmt}. To see how this works in both HQET and SCET let us introduce background $U(1)$ 2-form gauge fields $B\two_{e,m}$ such that:
\be
B\two_{e,m} \to B\two_{e,m} + d\Lambda\one_{e,m}\ec\qquad \int_{\Sigma_2} d\Lambda\one_{e,m} \in 2\pi \bZ\ed
\ee
When coupled to $B\two_{e,m}$, both effective actions \eqref{HQET_full} and \eqref{scet0} have terms at linear order in the background fields $B\two_{e,m}$ that read
\be
S_{\textrm{EFT}}|_{\cO(B)} = \int (B\two_e \wedge * J\two_e + B\two_m \wedge * J\two_m) \ed
\ee
In the presentation where one wants to preserve the electric 1-form symmetry, the second term above is entirely responsible for the mixed 't Hooft anomaly between $U(1)\one_e$ and $U(1)\one_m$. Namely, upon inspecting the full variation of the action:
\be
S_{\textrm{EFT}} [A\one + \Lambda\one_e, B\two_e + d\Lambda\one_e, B\two_m + d\Lambda\one_m] = S_{\textrm{EFT}} [A\one, B\two_e, B\two_m] + \frac{1}{2\pi}\int B\two_m \wedge d\Lambda\one_e\ec
\ee
we see that it fails to be invariant up to a c-number shift that only depends on background gauge fields.

The mixed anomaly underlies the Schwinger term in the Kac--Moody algebra which will be derived in section \ref{symmetriessoftthm}. In that language, it is simply the statement that electric and magnetic symmetry generators cannot be simultaneously represented without a central extension. Furthermore, we will use this central extension to fix the ordering-dependent contact term appearing in scattering amplitudes with soft particles of mixed electric--magnetic helicity.

\subsection{Scattering Amplitudes and Factorization}\label{factorization}
The effective theories HQET and SCET provide systematic descriptions of energetic or heavy particles interacting with soft gauge fields, making the separation of short- and long-distance dynamics manifest. This separation underlies the factorization theorems for scattering amplitudes \cite{Collins:1989gx}, where hard, collinear, and soft contributions can be organized and computed independently. See \cite{Agarwal:2021ais} for a  recent review of the subject. Typically, a scattering amplitude $\cA_n$ involving $n$ charged operators can be represented in the following way:
\begin{equation}
\label{eq:unified-factorization}
\,\mathcal{A}_n\!\left(p_i,m_a;\mu\right)
=
S\!\left(n_i,v_a;\mu\right)\,H\!\left(p_i,m_a;\mu\right)\,
\Bigg[\prod_{i\in \text{massless}} J_i\!\left(p_i;\mu\right)\Bigg]\,
+\;\text{power-suppressed}\ec
\end{equation}
where the indices $i$ and $a$ label respectively massless and massive legs such that $p^\mu_i = E_in^\mu_i$ and $p^\mu_a = m_a v_a^\mu$ with $n^2=0$ and $v^2 = 1$, $\mu$ is a renormalization scale and power corrections are suppressed by $k/m$ (HQET) or $\lambda$ (SCET). 

To unpack the above master formula, we now describe each factor in turn. First, the soft function $S(n_i,v_a;\mu)$, the object of primary importance in our discussion, is defined as the vacuum matrix element of Wilson lines:
\be
S(n_i,v_a;\mu)
=\Big\langle \prod_{i\in \text{massless}} Y_{Q_i}(n_i)
\prod_{a\in \text{massive}} W_{Q_a}(v_a)\Big\rangle\ec
\ee
with the lightlike Wilson lines $Y_{Q_i}(n_i)$ taken along the directions $n_i^\mu$ as in \eqref{xin} and the timelike lines $W_{Q_a}(v_a)$ along $v_a^\mu$ as in \eqref{newvar}.\footnote{This universal Wilson line sector is closely related to the asymptotic dressing that appears in infrared-finite formulations of the $S$-matrix see e.g. \cite{Kulish:1970ut,Hannesdottir:2019opa}. In that language, the same soft dynamics is encoded in dressed asymptotic states or, equivalently, in a finite hard $S$-matrix obtained by replacing the free Hamiltonian with a soft-collinear asymptotic Hamiltonian.} The factors $J_i(p_i;\mu)$, present only for massless charged legs, are the \emph{jet functions}: these are well-known SCET objects, they capture collinear dynamics and are closely related to the two-point correlator of gauge-invariant collinear fields in the same direction $n_i$. Their explicit form will not be needed below. Finally, the hard function $H(p_i,m_a;\mu)$ is IR finite and obtained by matching the full amplitude at the hard scale $\mu\sim\mathcal{Q}$ onto a local EFT operator inserted at the origin,
\be
H(p_i,m_a;\mu)=C(p_i,m_a;\mu)\,\langle \mathcal{O}_H\rangle\ec
\ee
where $C$ is a Wilson coefficient and $\mathcal{O}_H$ denotes the corresponding hard operator. As with the jet functions mentioned above, its detailed form is inessential for our purposes.
An immediate consequence of the formula \eqref{eq:unified-factorization} is that (to leading order in the expansion parameter) the scattering amplitude is essentially a product of Wilson lines
\be\label{prodwils}
\cA_n \simeq \left\langle \left[\prod_{i\in \text{massless}} Y_{Q_i}(n_i)\right]
\left[\prod_{a\in \text{massive}} W_{Q_a}(v_a)\right] \cO_H\right\rangle \times \Bigg[\prod_{i\in \text{massless}} J_i\!\left(p_i;\mu\right)\Bigg]\ed
\ee
The existence of conserved 1-form symmetry currents naturally raises the question: are there nontrivial $U(1)^{(1)}$ Ward identities (to be reviewed in section~\ref{symmetriessoftthm}) that the above scattering amplitude must satisfy? This question was answered positively by the authors of \cite{Berean-Dutcher:2025ohp} for massive charged particles using HQET. In addition, one can invoke the Goldstone theorem for spontaneously broken $U(1)^{(1)}$ symmetries~\cite{Gaiotto:2014kfa}, which states that
\be\label{1goldstone}
\langle\gamma_\varepsilon(q)|J_{\mu\nu}(x)\rangle = \frac{i}{e}(q_\mu\varepsilon_\nu^* - q_\nu\varepsilon^*_\mu)\ec
\ee
namely, the $U(1)^{(1)}$ current $J^{\mu\nu}$ has a nonzero overlap with a single-photon state of momentum $q$ and polarization $\varepsilon$, with the gauge coupling $e$ playing the role of a decay constant. Applying the $U(1)\one$ symmetry Ward identity to \eqref{prodwils} together with \eqref{1goldstone} implies
\be\label{softthm}
\cA_{n+\gamma_\varepsilon(q)} = e\sum^n_a Q_a \frac{\varepsilon \cdot v_a}{q \cdot v_a}\cA_n -i e q_\mu \varepsilon_\nu \langle 0|J^{\mu\nu}_H\prod^n_a W_{Q_a}(v_a)\cO_H|0\rangle\ed
\ee
In the strict soft limit $q\to 0$ the second term is at most $\mathcal{O}(q^0)$, and one thus recovers the universal soft-photon theorem; crucially, the correlator involving the hard part of the current cannot generate an additional $1/(q\!\cdot\!v)$ pole.

Some comments are in order:
\begin{itemize}
    \item Relative to \cite{Berean-Dutcher:2025ohp}, our analysis applies to both massive and massless charged particles, treated respectively within HQET and SCET. This also allows us to discuss, in a unified language, the role of the magnetic $U(1)^{(1)}_m$ symmetry and its consequences, which were not explored in the previous work.

    \item A second theme of this paper is the relation between higher-form symmetries and the discussion of soft theorems  in the asymptotic symmetry literature. In particular, in the next section we show that the electric and magnetic 1-form symmetries give rise to an infinite-dimensional algebra of ordinary $0$-form charges. This will allow us to recover the familiar asymptotic symmetry charges and their associated Ward identities from the higher-form symmetry perspective.

    \item We also explain how the same 1-form symmetry Ward identities can be extended beyond scattering amplitudes to inclusive observables. This is the natural framework for detector observables, and in particular for the photon detector discussed in section~\ref{sec:softsq-dglap-qed}, whose leading soft behavior can be understood from the same symmetry principles.
\end{itemize}

\section{From 1-form Symmetries to Asymptotic Symmetries and Soft Theorems}\label{symmetriessoftthm}

In this section we describe the relation between the electric and magnetic 1-form symmetries of the EFT and the familiar asymptotic-symmetry description of soft theorems. We also show how the electric-magnetic 1-form anomaly fixes a contact term appearing in scattering amplitudes involving two soft photons with mixed electric-magnetic polarizations.

\subsection{Abelian Kac-Moody Symmetry Enhancement}
In the previous section we showed that  both HQET and SCET possess electric and magnetic 1-form symmetries. In what follows we adapt the argument of \cite{Hofman:2018lfz} and show that from these 1-form  symmetries one can generate an infinite set of 0-form symmetries.

The basic mechanism is quite general. Given a tensor $T^{\mu_1 \dots \mu_\ell}$ satisfying a conservation equation
\be
\partial_{\mu_1} T^{\mu_1 \dots \mu_\ell}=0\,,
\ee
one can build a candidate 0-form symmetry current by contracting it with a parameter tensor $\epsilon$,
\be
J_\epsilon^\mu(x)=\epsilon_{\mu_2\dots \mu_\ell}(x)\,T^{\mu \mu_2 \dots \mu_\ell}(x)\,.
\ee
Requiring this current to be conserved imposes a differential condition on the parameter tensor:
\be
T^{\mu_1 \mu_2 \dots \mu_\ell}\,\partial_{\mu_1}\epsilon_{\mu_2\dots \mu_\ell}(x)=0\,,
\ee
whose precise form depends on the symmetry properties of $T$. For example, if $T$ is a symmetric tensor with $\ell=2$, as for the stress tensor, then $\epsilon_\mu$ must satisfy the Killing equation $\partial_\mu \epsilon_\nu+\partial_\nu \epsilon_\mu=0$ whose solutions are $\epsilon_\mu=a_\mu+b_{[\mu\nu]}x^\nu$, so the corresponding symmetry algebra is finite-dimensional with $a_\mu$ and $b_{[\mu\nu]}$ parametrizing translations and rotations. By contrast, when $T$ is an antisymmetric rank-two tensor, it is natural to write the associated conserved current in differential-form notation as
\[
j_\epsilon^{(1)}=* \big(T^{(2)}\wedge \epsilon^{(1)}\big)\,.
\]
Its conservation implies
\[
d\epsilon^{(1)}=0\,,
\]
namely that $\epsilon^{(1)}$ is a closed 1-form. While it may seem surprising that a 1-form symmetry can generate infinitely many ordinary symmetries, this is in fact quite natural: closed 1-forms are characterized by infinitely many degrees of freedom. For instance, all terms of the form $\epsilon^\mu\sim (x^2)^n x^\mu$ solve the closure condition independently for each $n$.\footnote{I would like to thank E. Trevisani for a discussion on this point.}

The case at hand is slightly richer since we have that a rank-two tensor and its Hodge dual are both conserved. 
It is then convenient to repackage the two 1-form symmetry currents $J\two_e$ and $J\two_m$ into a single complex-valued 2-form
\be
\cJ\two = \frac{1}{e^2}\left(F + i * F\right)\ec \qquad d\cJ\two = d*\cJ\two = 0~.
\ee
We further project the complex 2-form into its chiral/antichiral components,
\be
j\two \equiv \frac{1}{2\sqrt{\pi}}P \cJ\two\ec \qquad \bar{j}\two \equiv \frac{1}{2\sqrt{\pi}}\bar{P} ~\bar{\cJ}\two\ec
\ee
where $P$ and $\bar{P}$ are the following chiral/antichiral projectors
\be\label{proj}
P = \frac{1}{2} (1+i~*)\ec\qquad \bar{P} = \frac{1}{2} (1-i~*)\, .
\ee
By construction these currents satisfy $ d*j\two = d*\bar{j}\two = dj\two = d\bar{j}\two=0$.
Finally, by appropriately contracting these currents with a one-form parameter, we can define two chiral/anti-chiral  0-form symmetry currents 
\be
j\one_\Xi \equiv  * (j\two\wedge \bar{\Xi}\one ) \ec
\qquad
\bar{j}\one_{\bar{\Xi}} \equiv  * (\bar{j}\two\wedge \Xi\one)
\ed
\ee
Requiring the conservation of such currents, namely
 $d*j\one_\Xi=0$ and $d*\bar{j}\one_{\bar{\Xi}} =0$, we find that the 1-form parameters must satisfy 
 $\bar{P}~d\Xi\one =0$ and $P~d\bar{\Xi}\one=0$, indeed
 \be
0= d*j\one_\Xi =j\two\wedge d\bar{\Xi}\one =P j\two\wedge  d\bar{\Xi}\one=
 j\two\wedge  P d\bar{\Xi}\one 
 \Leftrightarrow  P d\bar{\Xi}\one =0  \, .
 \ee
 Then, for any oriented codimension-one hypersurface $\Sigma_3$, the charges
\be\label{KMcharges}
\bar{Q}_{\Xi}(\Sigma\three) \equiv \int_{\Sigma\three} * j\one_\Xi 
\, ,
\qquad
{Q}_{\bar{\Xi}}(\Sigma\three)  \equiv \int_{\Sigma\three} * \bar{j}\one_{\bar{\Xi}} 
\ec  
\ee
are conserved. These furnish an infinite-dimensional set of Abelian Kac--Moody-type charges; different choices of admissible $\Xi^{(1)}$ or $\bar\Xi^{(1)}$ label distinct symmetries. Using the canonical symplectic form of Maxwell theory and the conservation of $j^{(2)},\bar{j}^{(2)}$, one finds on any closed hypersurface $\Sigma^{(3)}$:
\be
\label{eq:KM-general}
\big[\,\bar Q_{\Xi_1}(\Sigma^{(3)}),\,\bar Q_{\Xi_2}(\Sigma^{(3)})\,\big]=0\,,\qquad
\big[\,Q_{\bar\Xi_1}(\Sigma^{(3)}),\,Q_{\bar\Xi_2}(\Sigma^{(3)})\,\big]=0\,,
\ee
while the mixed commutator is a pure c-number, commonly referred to as a Schwinger term:
\be
\label{eq:KM-mixed-general}
\big[\,\bar Q_{\Xi}(\Sigma^{(3)}),\,Q_{\bar\Xi}(\Sigma^{(3)})\,\big]
=\frac{1}{2\pi e^2}\,\int_{\Sigma^{(3)}} \Xi^{(1)}\wedge d\bar\Xi^{(1)}\,.
\ee

The upshot is that the emergent 1-form symmetries of HQET/SCET give rise to infinitely many conserved charges $\bar{Q}_\Xi,Q_{\bar{\Xi}}$ of an ordinary (albeit infinite-dimensional) zero-form symmetry. Notice that the charges \eqref{KMcharges} are non-trivial on compact surfaces $\Sigma\three$ only if  $\Xi\one,\bar{\Xi}\one$ are not exact. In the next section, we discuss how to relax this condition when considering Minkowski spacetime.

\subsection{Recovering Asymptotic Symmetry Charges}\label{sec:asymcharges}
Let us now specialize our discussion to Minkowski spacetime and allow the surface $\Sigma_3$ to run out to null infinity. Let $\epsilon(z,\bar z)$ be a function on the celestial sphere with antipodal matching at spatial infinity,\footnote{Our choice of local coordinates is specified in Appendix \ref{celcoord}.}
\be\label{antipmatching}
\epsilon(z,\bar z)\big|_{\mathcal I^+_-}=\epsilon(z,\bar z)\big|_{\mathcal I^-_+}\,.
\ee
We choose the chiral one-form near $\mathcal I^\pm$ such that
\be
\text{on }\mathcal I^+\!:\quad d\bar\Xi^{(1)} = d\epsilon(z,\bar z)\wedge du\,,
\qquad
\text{on }\mathcal I^-\!:\quad d\bar\Xi^{(1)} = -\,d\epsilon(z,\bar z)\wedge dv\,,
\ee
which satisfies $P\,d\bar\Xi^{(1)}=0$ on-shell. Using Stokes’ theorem and $dj^{(2)}=0$, the charge $Q_{\bar\Xi}$ reduces to a boundary integral. Taking the real (electric) part gives rise to the following conserved charges:
\be\label{eq:asym-charges-I}
Q^{\pm}_\epsilon \;=\; \frac{1}{e^2}\int_{\mathcal I^\pm} d\epsilon(z,\bar z)\wedge {*F}\,,
\ee
equivalently,
\be\label{eq:asym-charges-II}
Q^+_{\epsilon} = \frac{1}{e^2}\int_{\mathcal{I}^{+}_{-}} \epsilon(z,\bar{z})\, {*F}\,,\qquad
Q^-_{\epsilon} = \frac{1}{e^2}\int_{\mathcal{I}^{-}_{+}} \epsilon(z,\bar{z})\, {*F}\,.
\ee
Thus the charges defined in \eqref{KMcharges}, with the above choice of chiral data, reproduce the well-known asymptotic symmetry charges on $\mathcal I^\pm$ described in \cite{Strominger:2017zoo}.

Let us make some comments regarding this result:
\begin{itemize}
    \item The charges \eqref{KMcharges} may be defined on \emph{any} smooth oriented codimension-one hypersurface $\Sigma_3$, not only on $\mathcal I^\pm$. In particular, it will be useful to choose $\Sigma_3$ such that it includes pieces on $\mathcal I^\pm$ and, when desired, a small cap near timelike infinity $i^+$. These details will be important when we discuss the effects of Wilson lines piercing $\Sigma_3$ as in section \ref{softphotontheorem}.
    \item The same construction with $F$ in place of ${*}F$ yields the magnetic family of asymptotic symmetry charges discussed in \cite{Strominger:2015bla}:
\be
\tilde Q^{\pm}_{\tilde\epsilon}
= \frac{1}{2\pi}\int_{\mathcal I^\pm} d\tilde\epsilon(z,\bar z)\wedge F
\,,\qquad
\tilde{\epsilon}(z,\bar{z})\big|_{\mathcal I^+_-}=\tilde\epsilon(z,\bar{z})\big|_{\mathcal I^-_+}\,,
\ee
equivalently,
\be\label{magcharges}
\tilde Q^{+}_{\tilde\epsilon}
\;=\; \frac{1}{2\pi}\int_{\mathcal I^+_-} \tilde\epsilon(z,\bar z)\, F
\,,\qquad
\tilde Q^{-}_{\tilde\epsilon}
\;=\; \frac{1}{2\pi}\int_{\mathcal I^-_+} \tilde\epsilon(z,\bar z)\, F\,.
\ee 

   \end{itemize}

\subsection{Recovering the Soft Photon Theorem}\label{softphotontheorem}
An important consequence derived from the existence of higher-form global symmetries is that there are novel Ward identities constraining
the correlation functions of local operators in the presence of charged extended objects. In our study, the existence of a $U(1)\one$ 1-form
symmetry with conserved current $J\two$ acting on charged line operators $L_{Q_i}(C_i)$ leads to
\be\label{1formalternative}
\left\langle d * J\two(x)\prod^n_i L_{Q_i}(C_i)\cO\right\rangle = 
\sum^n_i Q_i \delta^{(3)}(C_i) \left\langle \prod_i L_{Q_i}(C_i)\, \mathcal{O} \right\rangle\ec
\ee
where $\delta^{(3)}(C_i)$ is the Poincar\`e dual form of $C_i$.\footnote{If the worldline $C_i$ is parametrized by $y_i(s)$ an expression in local coordinates is given by $[\delta^{(3)}(C_i)]_{\mu\nu\rho} = \int^\infty_0 ds \epsilon_{\mu\nu\rho\sigma}\frac{dy^\sigma_i}{ds} \delta^{(4)}(x - y_i(s))$.} Note that $\delta^{(3)}(C_i)$ is a three-form with the property that for any 1-form $\eta\one$ we have:
\be
\int_{X^{(4)}} \eta\one \wedge \delta^{(3)}(C_i) = \int_{C_i} \eta\one\ed
\ee
Let us now consider a codimension 2 closed surface $\Sigma\two$ which is the boundary of an open three-dimensional manifold $D\three$ such that $\partial D\three = \Sigma\two$. An immediate consequence of \eqref{1formalternative} is that, upon integrating both sides of the equation on $D\three$, we obtain \cite{Gaiotto:2014kfa}:
\be\label{integratedWard}
\left\langle Q(\Sigma\two)\prod^n_i L_{Q_i}(C_i)\cO_H\right\rangle = 
\sum^n_i Q_i\,\mathrm{Link}(\Sigma\two, C_i) \left\langle \prod_i L_{Q_i}(C_i)\, \mathcal{O}_H \right\rangle\ec
\ee
where $\mathrm{Link}(\Sigma\two, C_i)$ is the linking number defined as
\be
\mathrm{Link}(\Sigma\two, C_i) \equiv \int_{D\three} \delta^{(3)}(C_i)\ed
\ee
Equations \eqref{1formalternative} and \eqref{integratedWard} are central for us because, using the HQET/SCET variables introduced in the previous section, the leading-order scattering amplitude \eqref{prodwils} factors into a product of Wilson lines and a local hard operator. 

Since we work in Minkowski spacetime, we integrate both sides of \eqref{1formalternative} over future null infinity $\cI^+$ with an angle–dependent smearing function $\epsilon(z,\bar z)$. To connect with the emergent symmetries of the effective theories, we decompose the 1-form current $J^{(2)}$ into a radiative piece and a ``hard" piece $J\two_H$:
$J\two = \frac{1}{e^2}F\two + J\two_H$. Integrating by parts on $\mathcal{I}^+$ then gives:
\be
\begin{split}
    \int_{\mathcal{I^+}}&\epsilon(z,\bar{z})\left\langle d * J\two(x)\prod^n_i L_{Q_i}(C_i)\cO_H\right\rangle =\\& = \left\langle\frac{1}{e^2}\int_{\mathcal{I}^+} d\epsilon(z,\bar{z})\wedge *F(x) \prod^n_i L_{Q_i}(C_i)\cO_H\right\rangle + \int_{\cI^+} d\epsilon(z,\bar{z})\wedge \langle *J\two_H\prod^n_i L_{Q_i}(C_i)\cO_H\rangle\ed
\end{split}
\ee
Our interest here is precisely to connect the above equation to the leading soft photon theorem \eqref{softthm}. We recognize the first term as the insertion of the standard asymptotic charge \eqref{eq:asym-charges-I}.
The second term contains the hard current $J\two_H$. As discussed below \eqref{softthm}, its contribution is at most $\cO(q^0)$ 
and therefore cannot modify the universal leading $\cO(q^{-1})$ soft
pole. For this reason, we will drop this term in what follows.\footnote{Stokes' theorem also produces terms
supported on the end cuts of $\cI^+$. These are not assumed to
vanish: for a massive particle, this term measures the asymptotic
Coulomb field sourced by the particle. Their contributions do not provide an
additional radiative photon insertion. Consequently, they do not affect the
coefficient of the leading soft pole considered here.}

On the right-hand side of \eqref{1formalternative} we integrate the Poincar\'e-dual three-form over $\cI^+$, which localizes on the intersection of each worldline $C_i$ with $\cI^+$:
\be\label{eq:RHS-generic}
\sum_{i=1}^n Q_i\int_{\cI^+}\!\epsilon(z,\bar z)\,\delta^{(3)}(C_i)
\;=\;\sum_{i=1}^n Q_i\,\epsilon(\zeta_i)\,,
\ee
where $\zeta_i$ denotes the point on the celestial sphere picked out by the asymptotic direction of $C_i$. For massless legs, $\zeta_i=(z_i,\bar z_i)$ is the piercing point of $\cI^+$. On the other hand, for massive outgoing legs that asymptote to timelike infinity $i^+$, one may equivalently evaluate the charge on a mixed hypersurface 
$\Sigma^+=\mathcal I^+\cup\mathcal H^+$
where $\mathcal H^+$ is the unit timelike hyperboloid describing future timelike infinity.\footnote{Using our convention $v^2=1$, we parametrize a future-directed
four-velocity as
\be
v^\mu=\left(\sqrt{1+\rho^2},\,\rho\,\hat{\mathbf v}\right)\ec
\qquad
\rho=|\mathbf v|\ec
\qquad
\hat{\mathbf v}=\frac{\mathbf v}{|\mathbf v|}\,.
\ee
The celestial sphere function $\epsilon(z,\bar z)$ must then be extended to a function $\epsilon_{\mathcal H}(v)$ on $\mathcal H^+$ \cite{Campiglia:2015qka}:
\be
\Delta_{\mathcal H}\epsilon_{\mathcal H}=0\ec
\qquad
\lim_{\rho\to\infty}
\epsilon_{\mathcal H}(\rho,\hat{\mathbf v})
=
\epsilon(\hat{\mathbf v})\qquad \epsilon_{\mathcal H}(v)
=
\int_{S^2}\frac{d\Omega_{\hat q}}{4\pi}\,
\frac{\epsilon(\hat q)}
{\left(v^0-\mathbf v\cdot\hat{\mathbf q}\right)^2}\,.
\ee}
In the following we use $\epsilon(\zeta)$ as shorthand for this combined
function, with $\zeta=(z,\bar z)\in\mathcal I^+$ for a massless line and
$\zeta=v\in\mathcal H^+$ for a massive line.

To summarize, the Ward identity on $\cI^+$ can now be expressed as follows
\be\label{eq:Ward-Iplus-generic}
\Big\langle Q^+_{\epsilon}\,\prod_{i=1}^n L_{Q_i}(C_i)\,\cO_H\Big\rangle
\;=\;\sum_{i=1}^n Q_i\,\epsilon(\zeta_i)\;
\Big\langle \prod_{i=1}^n L_{Q_i}(C_i)\,\cO_H\Big\rangle\,.
\ee

Let us now specialize the generic lines $L_{Q_i}(C_i)$ to the dressings appropriate to massless and massive external legs:
\be
L_{Q_i}(C_i)\;\to\;\begin{cases}
Y_{Q_i}(n_i) & \text{for a massless leg in direction } n_i\,,\\
W_{Q_a}(v_a) & \text{for a massive leg with velocity } v_a\,.
\end{cases}
\ee
Using the factorized definition of the amplitude \eqref{prodwils} and the neutrality of the jet factors $J_i(p_i;\mu)$ under $U(1)^{(1)}$ symmetries, \eqref{eq:Ward-Iplus-generic} becomes
\be\label{eq:Ward-Iplus-both-final}
\begin{split}
\Big\langle Q^+_{\epsilon}\,&
\prod_{i\in \text{massless}} Y_{Q_i}(n_i)\,
\prod_{a\in \text{massive}} W_{Q_a}(v_a)\,\cO_H\Big\rangle
=\\
&\Bigg(
\sum_{i\in \text{massless}} Q_i\,\epsilon(z_i,\bar z_i)
+\sum_{a\in \text{massive}} Q_a\,\epsilon_{\cH}(v_a)
\Bigg)
\Big\langle
\prod_{i} Y_{Q_i}(n_i)\,
\prod_{a} W_{Q_a}(v_a)\,\cO_H
\Big\rangle\ed
\end{split}
\ee
At this point it is convenient to choose a meromorphic smearing 
\be\label{smear}
\epsilon(z,\bar z)=\frac{1}{w-z}\ec
\ee
which isolates a simple-pole structure on the celestial sphere \cite{Strominger:2017zoo}. With this choice the equation \eqref{eq:Ward-Iplus-both-final} now reads
\be\label{eq:soft-charge-amp-both-final}
Q^+_{\epsilon}\,\cA_n \;=\;
\Bigg(
\sum_{i\in \text{massless}} \frac{Q_i}{w-z_i}
+\sum_{a\in \text{massive}} Q_a \frac{\varepsilon_+(w,\bar w)\cdot v_a}
{\hat q(w,\bar w)\cdot v_a}
\Bigg)\,\cA_n\;,
\ee
valid at leading power. Here $\hat q^\mu(w,\bar w)$ denotes the unit-energy null vector and $(w,\bar{w})$ is a point on the celestial sphere.
\be
\hat q^\mu(w,\bar w)
=
\frac{1}{1+w\bar w}
\Big(
1+w\bar w,\,
w+\bar w,\,
-i(w-\bar w),\,
1-w\bar w
\Big)\,.
\ee
We also use the following parametrization for the positive-helicity polarization
\be
\varepsilon_+^\mu(w,\bar w)
=
\frac{1}{1+w\bar w}
\Big(
\bar w,\,
1,\,
-i,\,
-\bar w
\Big)\ec
\qquad
\hat q\cdot\varepsilon_+=0\,.
\ee
Moreover, for a massless momentum with $p_i^\mu=E_i\hat q^\mu(z_i,\bar z_i)$,
\be
\frac{\varepsilon_+(w,\bar w)\cdot p_i}
     {\hat q(w,\bar w)\cdot p_i}
=
\frac{1}{w-z_i}\,.
\ee
An $n$-point scattering amplitude $\cA_n$ describing $m$ incoming particles $\{|\mathrm{{in}}\rangle\}^m_{i=1}$ scattering into $n-m$ outgoing particles 
$\{|\mathrm{{out}}\rangle\}^n_{i=m+1}$ is given by
\be\label{smatrix}
\cA_n = \langle \mathrm{out} | \cS | \mathrm{in}\rangle\ec
\ee
where $\cS$ is the scattering matrix. Repeating the above construction on $\cI^-$ to define $Q^-_{\epsilon}$ with antipodally matched $\epsilon$ we arrive at:
\be\label{eq:S-matrix-ward-both-final}
\begin{split}
\langle \mathrm{out}|&\,(Q^+_{\epsilon}\,\cS-\cS\,Q^-_{\epsilon})\,|\mathrm{in}\rangle
=\\
&\Bigg[
\sum_{i\in\text{massless,in}} \frac{Q_i^{\mathrm{in}}}{w-z_i^{\mathrm{in}}}
+\sum_{a\in\text{massive,in}} Q^{\rm in}_a \frac{\varepsilon_+(w,\bar w)\cdot v^{\rm in}_a}
{\hat q(w,\bar w)\cdot v^{\rm in}_a}
\\
&-\sum_{j\in\text{massless,out}} \frac{Q_j^{\mathrm{out}}}{w-z_j^{\mathrm{out}}}
-\sum_{b\in\text{massive,out}} Q^{\rm out}_b \frac{\varepsilon_+(w,\bar w)\cdot v^{\rm out}_b}
{\hat q(w,\bar w)\cdot v^{\rm out}_b}
\Bigg]\,
\langle \mathrm{out}|\cS|\mathrm{in}\rangle\,.
\end{split}
\ee

A few comments are in order:
\begin{itemize}

\item Equation~\eqref{eq:soft-charge-amp-both-final} makes the connection
with the textbook leading soft-photon theorem \eqref{softthm} manifest. Writing $q^\mu=\omega\hat q^\mu(w,\bar w)$ and using the one-photon matrix element \eqref{1goldstone}, one then recovers 
\be
\lim_{q\to 0}\cA_{n+\gamma_\varepsilon(q)}
=
e\sum_{\ell=1}^n
\eta_\ell Q_\ell\,
\frac{p_\ell\cdot\varepsilon(q)}
     {p_\ell\cdot q}\,
\cA_n
+\cO(q^0)\ec
\ee
where $\eta_\ell=\pm 1$ if $\ell$ is outgoing/incoming. This form applies uniformly to massive and massless hard legs.

\item The same Ward identity argument applies to the magnetic 1-form symmetry with current $J^{(2)}_{m}=\tfrac{1}{2\pi}{*}F$ and charges defined in \eqref{magcharges}, acting on ’t~Hooft lines. With our normalizations, one obtains the \emph{magnetic} soft theorem described in \cite{Strominger:2015bla}:
\be\label{eq:mag-soft}
\lim_{q\to 0}\cA_{n+\tilde\gamma_{\tilde\varepsilon}(q)}
=
\frac{2\pi}{e}
\sum_{\ell=1}^n
\eta_\ell G_\ell\,
\frac{p_\ell\cdot\tilde\varepsilon(q)}
     {p_\ell\cdot q}\,
\cA_n
+\cO(q^0)\ec
\ee
where $G_a = \int *J\two_m$ is the magnetic charge and  $\tilde\varepsilon_\mu(q)$ is the dual polarization $\tilde\varepsilon_\mu \equiv \tfrac{1}{2}\epsilon_{\mu\nu\rho\sigma}\varepsilon^\nu q^\rho k^\sigma/(q\!\cdot\!k)$ for any reference $k^\mu$.\footnote{The overall coefficient in \eqref{eq:mag-soft} follows from our convention $J^{(2)}_{m}=\tfrac{1}{2\pi}{*}F$ and the ``1-form Goldstone'' overlap $\langle\gamma_{\tilde\varepsilon}(q)|J_{m\,\mu\nu}(0)|0\rangle=\tfrac{ie}{2\pi}\!\left(q_\mu\tilde\varepsilon_\nu^*-q_\nu\tilde\varepsilon_\mu^*\right)$.}

\item Conceptually, the derivation above bridges the asymptotic symmetry description of soft theorems (see \cite{Strominger:2017zoo} and references therein) with the generalized symmetry viewpoint \cite{Gaiotto:2014kfa,Berean-Dutcher:2025ohp}. On the one hand, in the asymptotic-symmetry approach soft Ward identities are phrased as consequences of large gauge transformations of $A_\mu$; here we show that the same constraints follow from the \emph{emergent} one-form symmetries of the soft/EFT limit. On the other hand, standard discussions of generalized symmetries often focus on finite-dimensional conserved charges; adapting the construction of \cite{Hofman:2018lfz} to our EFTs yields an \emph{infinite-dimensional} Abelian Kac--Moody symmetry whose Ward identities reproduce the known soft theorems.

\end{itemize}

\subsection{A Consequence of the Mixed Anomaly}

In this subsection we show that, in QED, the soft limit of scattering amplitudes involving two soft photons with mixed
electric--magnetic polarizations contains a universal, ordering-dependent contact term. Such terms have also appeared in the context of celestial CFTs (see, e.g.~\cite{Nande:2017dba}), although they are often discarded via a normal-ordering prescription since
they correspond to disconnected diagrams. Here we show instead that there is a physical prescription for these contact terms
provided by the mixed 1-form symmetry anomaly.

Let us first consider the single soft photon theorems. For an $(n{+}1)$-point amplitude with one electric soft photon $(q,\varepsilon)$,
\be
\label{eq:single-soft-e}
\cA_{n+1}^{(e)}(q)\;=\;S_e(q)\,\cA_n\;+\;\cO(q^0),
\qquad
S_e(q)=e\sum_{i=1}^n Q_i\,\frac{p_i\!\cdot\!\varepsilon}{p_i\!\cdot\! q}\,,
\ee
and similarly for a magnetic soft insertion (dual polarization $\tilde\varepsilon_\mu$),
\be
\label{eq:single-soft-m}
\cA_{n+1}^{(m)}(q)\;=\;S_m(q)\,\cA_n\;+\;\cO(q^0),
\qquad
S_m(q)=\frac{2\pi}{e}\sum_{i=1}^n G_i\,\frac{p_i\!\cdot\!\tilde\varepsilon}{p_i\!\cdot\! q}\,.
\ee
Let us now consider the $(n{+}2)$-point amplitude with one electric soft leg $(q_1,\varepsilon_1)$ and one magnetic soft leg $(q_2,\tilde\varepsilon_2)$. In the \emph{ordered} soft limit $q_1\to0$ followed by $q_2\to0$, the leading structure is constrained to 
\be
\label{eq:double-soft-ansatz}
\cA_{n+2}^{(e\to m)}(q_1,q_2)
\;=\;S_e(q_1)\,S_m(q_2)\,\cA_n\;+\;R(q_1,q_2)\,\cA_n\;+\;\cO(q_1^0,q_2^0)\,,
\ee
where $R$ is a scalar distribution in the soft variables such that:
\begin{itemize}
    \item It is gauge invariant in each leg.
    \item It carries no extra poles in $p_i\!\cdot\! q_{1,2}$.
    \item It has support only at coincident insertion points on the celestial sphere.
\end{itemize}
The only solution to the above requirements at leading order is a contact term on the celestial sphere,
\be
R(q_1,q_2) = K \varepsilon_1 \cdot \tilde{\varepsilon}_2~\delta^{(2)}(\hat{q}_1 - \hat{q}_2)~,
\ee
where $\hat{q}$ denotes the direction of the soft momentum on the celestial sphere and $K$ is a universal coefficient fixed below.\footnote{Derivatives of $\delta^{(2)}$ would require additional angular derivatives (equivalently extra factors of the soft momenta) and hence
enter only at subleading order in the soft expansion.
} The derivation of the polarization factor $\varepsilon_1 \cdot \tilde{\varepsilon}_2$, which is gauge invariant on the support of the delta function, is discussed in appendix \ref{sec:freefield}.

Now, to fix the coefficient $K$ we can proceed as follows. From the discussion in section \ref{symmetriessoftthm}, we recall that the mixed commutator between Kac-Moody charges in \eqref{eq:KM-mixed-general} specialized to future null infinity $\cI^+$ leads to
\be
[Q_e(\epsilon_1),Q_m(\tilde{\epsilon}_2)] = \frac{1}{2\pi e^2}\int_{S^2} d\epsilon_1 \wedge d\tilde{\epsilon}_2\ed
\ee
In the meromorphic basis where $\epsilon_1(z,\bar z)=1/(z-w_1)$, $\tilde\epsilon_2(z,\bar z)=1/(\bar z-\bar w_2)$ one obtains the local form of the central extension
\be
\label{eq:KM-local}
\big[\,Q_e(w_1),\,Q_m(w_2)\,\big]=\frac{2\pi}{e^2}\,\delta^{(2)}(w_1-w_2)\ed
\ee
A different derivation of the above central extension, using free fields, is reported in appendix \ref{sec:freefield}. 

Acting on the S-matrix \eqref{smatrix} with the above and using LSZ, we notice that the antisymmetrized ordered double-soft limit must reproduce the c-number appearing in \eqref{eq:KM-local}:
\be
\label{eq:double-soft-central}
\Big[\cA_{n+2}^{(e\to m)}-\cA_{n+2}^{(m\to e)}\Big](q_1,q_2)
\;=\;\frac{2\pi}{e^2}\, \varepsilon_1\!\cdot\!\tilde\varepsilon_2  \,\delta^{(2)}\!\big(\hat q_1-\hat q_2\big)\,\cA_n
\;+\;\cO(q_1^0,q_2^0)\,.
\ee
As a result, our analysis fixes the coefficient to be $K=\tfrac{2\pi}{e^2}$. The complete leading mixed double-soft theorem is then given by
\be
\label{eq:double-soft-final}
\cA_{n+2}^{(e\to m)}(q_1,q_2)
\;=\;S_e(q_1)\,S_m(q_2)\,\cA_n
\;+\;\frac{2\pi}{e^2}\varepsilon_1 \cdot \tilde{\varepsilon}_2\,\delta^{(2)}\!\big(\hat q_1-\hat q_2\big)\,\cA_n
\;+\;\cO(q_1^0,q_2^0)\ed
\ee

A few remarks are in order. First, the above contact term is universal and independent of the hard process: it is completely fixed by the mixed electric-magnetic 1-form anomaly. In this sense, the ordering dependence of the mixed double-soft limit is not an ambiguity of the soft expansion, but a physical consequence of the fact that electric and magnetic soft charges do not commute. Second, the support of the contact term at coincident celestial angles is the momentum-space counterpart of the local central extension in the Kac--Moody algebra on the celestial sphere. This explains why such terms are often absent in treatments based on normal ordering: away from coincident insertions the anomaly is invisible, but once one keeps track of the local operator algebra, the contact term is forced by the anomaly.

\section{1\text{-}Form Symmetry Constraints on Inclusive Observables}
\label{sec:softsq-dglap-qed}

In this section we extend the 1-form symmetry derivation of the soft-photon theorem for scattering amplitudes in section \ref{symmetriessoftthm} to an observable-level statement.
The key new ingredient is that inclusive observables are naturally formulated as in--in (``cut'')
correlators, so the same asymptotic charge $Q_\epsilon^\pm$ acts on both sides of the cut. Similar ideas, formulated in the language of asymptotic symmetries, have been recently used in \cite{Gonzalez:2025ene,Moult:2025njc} to discuss soft theorems for inclusive observables in gravity. Finally,  we also connect this symmetry statement
to light-ray operators and DGLAP detectors recently discussed in \cite{Chang:2025zib}.

\subsection{Inclusive Observables, Wilson-line Factorization and Ward Identity}
\label{subsec:cut-wilson-factor}

We work to leading power in the HQET/SCET expansion parameters so that scattering amplitudes factorize
into Wilson-line dressings times a hard insertion, as in section \ref{factorization}.

To discuss cross sections and detector insertions, we consider the in--in (cut) object
\be
\label{eq:inin-master}
\langle \mathcal M\rangle_{\rm in}
\;\equiv\;
\langle \mathrm{in}|\,\cS^\dagger\,\mathcal M\,\cS\,|\mathrm{in}\rangle\,,
\ee
where $\mathcal M$ is any measurement operator acting on the out Hilbert space.
By the discussion of section \ref{factorization}, at leading power, the soft sector of $\cS$ is captured by the Wilson-line operator
$\mathcal O_{\rm soft}$, hence the soft contribution to \eqref{eq:inin-master} factorizes as
\be
\label{eq:inin-soft-factorized}
\langle \mathcal M\rangle_{\rm in}
\;\simeq\;
\langle 0|\,
\mathcal O_H^\dagger\,
\mathcal O_{\rm soft}^\dagger\,
\mathcal M\,
\mathcal O_{\rm soft}\,
\mathcal O_H
\,|0\rangle
\times(\text{jets/hard matching})\,,
\ee
where ``jets/hard matching'' denotes the usual collinear factors and hard Wilson coefficients that do not
participate in the soft Ward identities. Equation \eqref{eq:inin-soft-factorized} is the cut analogue of
\eqref{prodwils}: it makes manifest that soft physics in inclusive observables is governed by
correlators of Wilson lines, now with an insertion $\mathcal M$ on the cut.

Let us now recall the asymptotic electric charges from section \ref{sec:asymcharges}:
\be
Q^{+}_\epsilon \;=\; \frac{1}{e^2}\int_{\mathcal I^+} d\epsilon(z,\bar z)\wedge {*F}\,,
\qquad
Q^{-}_\epsilon \;=\; \frac{1}{e^2}\int_{\mathcal I^-} d\epsilon(z,\bar z)\wedge {*F}\,,
\ee
with antipodal matching at spatial infinity as in \eqref{antipmatching}.
Because $Q_\epsilon$ implements a global symmetry, its action on any operator is by commutator.
Applying this to the inclusive observable \eqref{eq:inin-master} yields the in-in Ward identity
\be
\label{eq:cut-ward-identity}
\langle \mathrm{in}|\,\cS^\dagger\,[Q_\epsilon^+,\mathcal M]\,\cS\,|\mathrm{in}\rangle
\;=\;
\langle \mathrm{in}|\Big(\cS^\dagger Q_\epsilon^+\,\mathcal M\,\cS
-\cS^\dagger\,\mathcal M\,\cS\,Q_\epsilon^-\Big)|\mathrm{in}\rangle\,.
\ee
At leading power we may replace the soft part of $\cS$ in \eqref{eq:cut-ward-identity} by
$\mathcal O_{\rm soft}$, yielding the Wilson-line EFT version
\be
\label{eq:cut-ward-wilson}
\Big\langle 0\Big|\,
\mathcal O_H^\dagger\,
\Big(
\mathcal O_{\rm soft}^\dagger\,[Q_\epsilon^+,\mathcal M]\,\mathcal O_{\rm soft}
\Big)\,
\mathcal O_H
\Big|0\Big\rangle
=
\Big\langle 0\Big|\,
\mathcal O_H^\dagger\,
\Big(\mathcal O_{\rm soft}^\dagger Q_\epsilon^+\,\,\mathcal M\,\mathcal O_{\rm soft}
-\mathcal O_{\rm soft}^\dagger\,\mathcal M\,\mathcal O_{\rm soft}\,Q_\epsilon^-
\Big)\,
\mathcal O_H
\Big|0\Big\rangle\,,
\ee
where, as in section~\ref{factorization}, the hard operator $\mathcal O_H$ does not contribute soft poles.

\subsection{Cross-Section Soft Theorem from the In-in Ward Identity}
\label{subsec:soft-squared}

As in section \ref{softphotontheorem}, let us parametrize a null momentum by an energy $\omega>0$ and a point $(w,\bar{w})$:
\be
\label{eq:celestial-q}
q^\mu(\omega;w,\bar w)
=
\omega\,
\hat q^\mu(w,\bar w)\,,
\qquad
\hat q^\mu(w,\bar w)
=
\frac{1}{1+w\bar w}\Big(1+w\bar w,\; w+\bar w,\;-i(w-\bar w),\;1-w\bar w\Big)\,,
\ee
so that $q^2=0$ and $\omega=q^0$.
A convenient choice of polarization vectors is the standard helicity basis
$\varepsilon^\mu_\pm(w,\bar w)$ satisfying $q\cdot\varepsilon_\pm=0$ and
$\varepsilon_\pm\cdot\varepsilon_\pm=0$.

We further define the mode-resolved number operator for an outgoing photon of energy $\omega$ and direction $(w,\bar w)$:
\be
\label{eq:number-operator-celestial}
\mathcal N(\omega;w,\bar w)
\;\equiv\;
\sum_{h=\pm}\,
a^\dagger_{h}(\omega;w,\bar w)\,a_{h}(\omega;w,\bar w)\,.
\ee
Our goal is to study the implications of the inclusive Ward identity \eqref{eq:cut-ward-identity} when we set the measurement operator $\mathcal M$ to be the insertion of a single
outgoing photon in a fixed mode, $\mathcal M=\mathcal N(\omega;w,\bar w)$.

Notice that the charge $Q_\epsilon^+$ is linear in the radiative Maxwell field on $\mathcal I^+$, hence linear in the oscillators
$a_h,a_h^{\dagger}$. Therefore the in-in Ward
identity inserts a single soft photon on one side of the cut. The emission probability is therefore obtained by
combining the insertion on the ket and on the bra. Concretely, choosing the same meromorphic smearing as in \eqref{smear}, the Wilson line Ward identity derived in section~\ref{softphotontheorem} implies that $Q^+_\epsilon$ acts on the soft
operator $\mathcal O_{\rm soft}$ by multiplication:
\be
\label{eq:charge-action-soft}
Q^+_{\epsilon}\,\mathcal O_{\rm soft}
=
\Bigg(
\sum_{i\in \text{massless}} \frac{Q_i}{w-z_i}
+\sum_{a\in \text{massive}} Q_a \frac{\varepsilon_+(w,\bar w)\cdot v_a}
{\hat q(w,\bar w)\cdot v_a}
\Bigg)\,\mathcal O_{\rm soft}\,,
\ee
and similarly for the action of $Q^-_\epsilon$ on incoming lines. Inserting this action into the  correlator
\eqref{eq:inin-soft-factorized} and applying \eqref{eq:cut-ward-wilson} yields, at leading power,
\be
\label{eq:soft-squared-celestial}
\Big\langle \mathcal N(\omega;w,\bar w)\Big\rangle_{\rm in}
\;\xrightarrow[\omega\to 0]{}\;
\Bigg|
\sum_{i\in \text{massless}} \frac{e\,Q_i}{w-z_i}
+\sum_{a\in \text{massive}}eQ_a\frac{\varepsilon_+(w,\bar w)\cdot v_a}
{\hat q(w,\bar w)\cdot v_a}
\Bigg|^2
\Big\langle \mathbf{1}\Big\rangle_{\rm in}
\,+\,\cO(\omega^{-1})\,,
\ee
where $\langle \mathbf 1\rangle_{\rm in}$ denotes the corresponding $n$-point inclusive observable without the measured
soft photon.
Equation \eqref{eq:soft-squared-celestial} is the celestial sphere representation of the soft theorem for cross sections:
it is literally the product of the holomorphic and antiholomorphic Ward identity factors, one from each side of
the cut.

To recover the standard momentum-space form, note that the meromorphic poles map to the eikonal factors
\be
\label{eq:celestial-to-eikonal}
\frac{1}{w-\zeta_a}
\quad\longleftrightarrow\quad
\frac{p_a\cdot \varepsilon_+(q)}{p_a\cdot q}\,,
\qquad
\frac{1}{\bar w-\bar\zeta_a}
\quad\longleftrightarrow\quad
\frac{p_a\cdot \varepsilon_-(q)}{p_a\cdot q}\,,
\ee
where $\zeta_a$ denotes the celestial direction associated to a given leg $a$. Thus \eqref{eq:soft-squared-celestial} is equivalent to the usual factorization of the soft photon
cross-section:
\be
\label{eq:soft-squared-master}
d\sigma_{n+\gamma(q)} \;\xrightarrow[q\to 0]{}\;
\sum_{\rm pol}\Big|S^{(0)}(q)\Big|^2\,
d\sigma_n\,
\frac{d^3\vec q}{(2\pi)^3\,2\omega}
\;+\;\cO(q^0)\,,
\ee
where
\be\label{eikonalKernel}
S^{(0)}(q)\;=\; e\sum_{a=1}^n Q_a\,\frac{p_a\!\cdot\!\varepsilon(q)}{p_a\!\cdot\! q}\,,
\qquad
\sum_{\rm pol}\Big|S^{(0)}(q)\Big|^2
=
e^2\sum_{a,b=1}^n Q_aQ_b\,
\frac{p_a\!\cdot\! p_b}{(p_a\!\cdot\! q)(p_b\!\cdot\! q)}\,.
\ee
This analysis shows that the universal soft divergence for cross
sections is fixed by symmetry, just as it is for amplitudes, provided
one uses the correct in--in Ward identity
\eqref{eq:cut-ward-identity}.

\subsection{DGLAP Detectors for QED}

In the previous section we derived the leading cross-section soft factor from an in-in Ward identity for the same 1-form symmetry charge $Q_\epsilon^\pm$ that produced the amplitude soft theorem in section ~\ref{symmetriessoftthm}. We now explain how this symmetry derivation connects with the analysis of DGLAP detectors in \cite{Chang:2025zib}, which relates the location of poles in the detector boost weight $J_L$ to the soft theorem.

The DGLAP detectors discussed in the context of QCD in \cite{Chang:2025zib} are a generalization of the energy flow operator in perturbation theory. Their definition, adapted to QED, is as follows:
\be
\label{eq:dglap-detector-def}
\mathcal D_{J_L}(w,\bar w)
\;\equiv\;
\sum_{h=\pm}\int_0^\infty
\frac{d\omega}{(2\pi)^{d-1}\,2\omega}\;
\omega^{-J_L}\;
a^\dagger_{h}\!\big(\omega; w,\bar w\big)\,
a_{h}\!\big(\omega; w,\bar w\big)~.
\ee
Their role is to measure photons at a fixed point $(w,\bar w)$ on the celestial sphere and to assign a weight $\omega^{2-d-J_L}$. The complex parameter $J_L$ is interpreted as a boost weight along a null direction which labels irreducible representations of the Lorentz group $SO(d-1,1)$.

Let us now consider the inclusive observable with the detector insertion,
\be
\label{eq:detector-inin}
\langle \mathcal D_{J_L}(w,\bar w)\rangle_{\rm in}
\;=\;
\langle \mathrm{in}|\cS^\dagger\,\mathcal D_{J_L}(w,\bar w)\,\cS|\mathrm{in}\rangle\,.
\ee
Using the definition \eqref{eq:dglap-detector-def} it follows that:
\be
\label{eq:dglap-onept-start}
\langle \mathcal D_{J_L}(w,\bar w)\rangle_{\rm in}
=
\int_0^\infty\frac{d\omega}{(2\pi)^{d-1}\,2\omega}\;
\omega^{-J_L}\;
\langle \mathcal{N}(\omega;w,\bar w)\rangle_{\rm in}~,
\ee
with $\cN(\omega, w,\bar{w})$ the photon number operator defined in \eqref{eq:number-operator-celestial}. In the soft limit $q^\mu=\omega\,\hat{q}^\mu(w,\bar w)$ with $\omega\to0$, the in-in Ward identity fixes the leading
behavior of $\cN(\omega, w,\bar{w})$ to be the usual squared eikonal kernel \eqref{eikonalKernel},
\be
\label{eq:soft-squared-omega-factor}
e^2\sum_{a,b}Q_aQ_b\,\frac{p_a\!\cdot p_b}{(p_a\!\cdot q)(p_b\!\cdot q)}
=
\frac{1}{\omega^2}\,\mathcal K(w,\bar w;\{p\}),
\qquad
\mathcal K(w,\bar w;\{p\})\equiv
e^2\sum_{a,b}Q_aQ_b\,\frac{p_a\!\cdot p_b}{(p_a\!\cdot \hat{q})(p_b\!\cdot \hat{q})}\,,
\ee
Inserting \eqref{eq:soft-squared-omega-factor} into \eqref{eq:dglap-onept-start}, the soft region contributes
\be
\label{eq:JL-pole-integrand}
\langle \mathcal D_{J_L}(w,\bar w)\rangle_{\rm in}
\;\supset\;\frac{\mathcal K(w,\bar w;\{p\})}{2(2\pi)^{d-1}}\,
\langle \mathbf 1\rangle_{\rm in}\,
\int_0^\delta d\omega\;\omega^{-J_L-3}
\;+\;\cdots\,,
\ee
where $\delta$ is an effective upper cutoff, set by the breakdown of the strict soft approximation, as in
\cite{Chang:2025zib}, and the ellipses denote terms less singular as $\omega\to0$. It thus follows that
\be
\label{eq:JL-pole-final}
\langle \mathcal D_{J_L}^{(\gamma)}(w,\bar w)\rangle_{\rm in}
\;\supset\;
-\frac{1}{J_L+2}\;
\frac{\mathcal K(w,\bar w;\{p\})}{2(2\pi)^{d-1}}\,
\langle \mathbf 1\rangle_{\rm in}
\;+\;\text{regular}\,.
\ee
The DGLAP photon detector has a universal simple pole at $J_L=2-d=-2$ (in $d=4$ for spin-1)\footnote{In contrast to a finite-resolution measurement, the idealized
detector \eqref{eq:dglap-detector-def} is not itself IR safe in QED, since a sharp photon-number
observable is sensitive to arbitrarily soft radiation. A physically finite detector would instead require an explicit energy and angular resolution. In that case the lower endpoint of the detector integral is set by the detector threshold rather than $\omega=0$, and the sharp pole is replaced by dependence on the detector resolution scale, reducing to a logarithm at $J_L=-2$.}, whose residue is fixed entirely by the same leading soft theorem obtained from the in-in Ward identity.

Unlike in QCD, where the detector spectrum near $J_L=2-d$ is affected by DGLAP/BFKL trajectory\footnote{BFKL stands for Balitsky-Fadin-Kuraev-Lipatov.} recombination discussed in \cite{Chang:2025zib}, we do not expect a similar mechanism to take place. The reason is that there is no direct QED analogue of the BFKL detector associated with gluon reggeization. Hence the pole at $J_L=-2$ found here is simply interpreted as the Mellin transform of the universal soft-photon behavior, not as a signal of trajectory recombination.

\section*{Acknowledgements}\noindent
I am particularly grateful to Luc\'ia C\'ordova, Shota Komatsu, and Emilio Trevisani for discussions and feedback at various stages of this project. I also thank Gabriel Cuomo, Andreas Helset, Diego Hofman, Barak Gabai, Julio Parra-Martinez and Sasha Zhiboedov for useful comments and discussions.

\appendix

\section{Bondi Charts}\label{celcoord}
In the asymptotic symmetry literature, a popular choice of coordinates is the retarded Bondi chart $x^\mu=(u,r,z,\bar z)$, in which the flat Minkowski metric (with $\eta_{ab}$ Cartesian) takes the form
\be
\label{Bondimetric}
ds^2 = -dudr + r^2dzd\bar{z}\ed
\ee
Following \cite{Dumitrescu:2015fej}, the explicit map from Bondi to Cartesian $y^a$ is
\be
\label{bondicoord}
y^0 = \tfrac12\!\left(u + r(1+|z|^2)\right),\quad
y^1 = \tfrac{r}{2}(z+\bar z),\quad
y^2 = -\,\tfrac{i r}{2}(z-\bar z),\quad
y^3 = -\,\tfrac12\!\left(u - r(1-|z|^2)\right)\,.
\ee
Here $u,r\in \mathbb{R}$, $z\in\mathbb C$. Future null infinity $\mathscr I^+$ is located at $r\to\infty$ with topology $\mathbb R_u\times S^2$, and its future/past boundaries $\mathscr I^+_\pm$ sit at $u\to\pm\infty$. 

\paragraph{Advanced (ingoing) Bondi chart near $\mathscr I^-$.}
For completeness, we also record the advanced Bondi chart $x^\mu=(v,r,z,\bar z)$, adapted to $\mathscr I^-$. In this chart, the metric reads
\be
\label{Bondi-advanced-metric}
ds^2 = -dvdr + r^2dzd\bar{z}\ed
\ee
The map to Cartesian coordinates is obtained from \eqref{bondicoord} by replacing $u\to v$ and exchanging the retarded with the advanced null direction,\footnote{We keep the same complex stereographic coordinate $z$ on the celestial $S^2$ for both charts; antipodal matching across spatial infinity imposes $\epsilon(z,\bar z)|_{\mathscr I^+_-}=\epsilon(z,\bar z)|_{\mathscr I^-_+}$.}
\be
\label{bondicoord-advanced}
y^0 = \tfrac12\!\left(v + r(1+|z|^2)\right),\quad
y^1 = \tfrac{r}{2}(z+\bar z),\quad
y^2 = -\,\tfrac{i r}{2}(z-\bar z),\quad
y^3 = \;\tfrac12\!\left(v - r(1-|z|^2)\right)\,,
\ee
so that past null infinity $\mathscr I^-$ is again at $r\to\infty$ with null coordinate $v$ and celestial sphere $(z,\bar z)$.

\section{Free-Field Derivation of the Mixed Anomaly Contact Term}\label{sec:freefield}
In the main text we found the relation \eqref{eq:KM-local}. Let us see how such a contribution  is related to a disconnected diagram in $4d$ Maxwell theory.\footnote{I would like to thank E. Trevisani for suggesting the inclusion of this appendix.} 
To start, we expand the gauge field $A^\mu$ in modes
\begin{equation}
    A^\mu(x)
=
e\sum_{h=\pm1}\int\widetilde{d^3k}\,
\left[
\varepsilon_h^\mu(k)a_h(\vec k)e^{-ik\cdot x}
+\text{h.c.}
\right]~,
\end{equation}
where
\begin{equation}
\widetilde{d^3k}\equiv
\frac{d^3\vec k}{(2\pi)^3\,2\omega_{\vec k}}~,
\qquad \omega_{\vec{k}}=|\vec{k}|~,\qquad
[a_h(\vec k),a_{h'}^\dagger(\vec k\,')]
=(2\pi)^3\,2\omega_{\vec k}\,
\delta_{hh'}\delta^{(3)}(\vec k-\vec k\,')~.
\end{equation}
and $h$ is the helicity. The corresponding expansions of $F^{\mu\nu}$ and
$\tilde F^{\mu\nu}\equiv
\frac{1}{2e^2}\epsilon^{\mu\nu\rho\sigma}F_{\rho\sigma}$ are
\begin{align}
F^{\mu\nu}(x)
&=
e\sum_{h=\pm1}\int\widetilde{d^3k}\,
\left[
-i f_h^{\mu\nu}(k)a_h(\vec k)e^{-ik\cdot x}
+\text{h.c.}
\right],
\\
\tilde F^{\mu\nu}(x)
&=
e\sum_{h=\pm1}\int\widetilde{d^3k}\,
\left[
-i\tilde f_h^{\mu\nu}(k)a_h(\vec k)e^{-ik\cdot x}
+\text{h.c.}
\right].
\end{align}
where
\begin{equation}
f_h^{\mu\nu}(k)
=
k^\mu\varepsilon_h^\nu(k)
-k^\nu\varepsilon_h^\mu(k)\,.
\end{equation}
We choose the Levi--Civita convention for which Hodge duality acts on
the helicity modes according to
\begin{equation}
\label{tildef}
\tilde f_h^{\mu\nu}(k)
\equiv
\frac{1}{2e^2}
\epsilon^{\mu\nu}{}_{\rho\sigma}
f_h^{\rho\sigma}(k)
=
-\frac{ih}{e^2}f_h^{\mu\nu}(k)
=
\frac{1}{e^2}
\left(
k^\mu\tilde\varepsilon_h^\nu(k)
-k^\nu\tilde\varepsilon_h^\mu(k)
\right),
\qquad
\tilde\varepsilon_h^\mu(k)
\equiv
-ih\,\varepsilon_h^\mu(k)\,.
\end{equation}
Thus, for each helicity mode, $\tilde F^{\mu\nu}$ differs from
$F^{\mu\nu}$ by the factor $-ih/e^2$. A potential for $\tilde F$
may be introduced, up to a gauge transformation, as
\be
\tilde A^\mu(x)
=
e\sum_{h=\pm1}\int\widetilde{d^3k}\,
\left[
\varepsilon_h^\mu(k)\tilde a_h(\vec k)e^{-ik\cdot x}
+\text{h.c.}
\right]\ed
\ee
Therefore, the ordinary one-photon state and the state created by the dual-field insertion are, respectively,
\begin{equation}
    |\gamma_h(\vec{k})\rangle\equiv a_h^\dagger(\vec{k})|0\rangle
    \, ,
    \qquad 
       |\tilde\gamma_h(\vec k)\rangle_{\mathrm{ins}}
\equiv
\tilde a_h^\dagger(\vec k)|0\rangle
=
\frac{ih}{e^2}\,
a_h^\dagger(\vec k)|0\rangle
=
\frac{ih}{e^2}\,
|\gamma_h(\vec k)\rangle\,.
\end{equation}
The latter denotes a state created by the dual-field insertion and is not an independently normalized asymptotic photon state. In the antisymmetrized product of the two soft insertions, the mixed contraction contains the disconnected piece
\begin{equation}
\label{eq:disconnected-mixed-contraction}
\langle 0|
\widetilde a_{h'}(\vec k\,')
a_h^\dagger(\vec k)
|0\rangle
=
-\frac{ih'}{e^2}
(2\pi)^3\,2\omega_{\vec k}\,
\delta^{(3)}(\vec k\,'-\vec k)\,
\delta_{h,h'}\,.
\end{equation}
After projecting onto the soft modes and antisymmetrizing the two orderings, this disconnected contribution gives the angular contact term in \eqref{eq:KM-local}.

Equation \eqref{eq:disconnected-mixed-contraction} is diagonal in
helicity. Since Hodge duality acts on a helicity-$h$ polarization as
$\tilde{\varepsilon}_h=-ih\,\varepsilon_h$, the factor $-ih/e^2$
is the duality phase multiplied by the normalization $1/e^2$.
Consequently, the disconnected contraction has the mixed polarization
structure $\varepsilon_1\cdot\tilde{\varepsilon}_2$ appearing in the
main text. On the support of the delta function,
$\vec k\,'=\vec k$, and this pairing is gauge invariant. Indeed,
\begin{equation}
(\varepsilon_1+\alpha k)\cdot
(\tilde{\varepsilon}_2+\beta k)
=
\varepsilon_1\cdot\tilde{\varepsilon}_2\,,
\end{equation}
where we used
$k\cdot\varepsilon_1
=k\cdot\tilde{\varepsilon}_2=k^2=0$.

\bibliographystyle{ytphys}
\baselineskip=0.85\baselineskip
\bibliography{detectors}

\end{document}